\documentclass[12pt]{article}

\usepackage{scicite}

\usepackage{times}

\usepackage{graphicx}
\usepackage[space]{grffile}
\usepackage{latexsym}
\usepackage{amsfonts,amsmath,amssymb}
\usepackage{url}
\usepackage[utf8]{inputenc}
\usepackage{hyperref}
\hypersetup{colorlinks=false,pdfborder={0 0 0}}
\usepackage{textcomp}
\usepackage{longtable}
\usepackage{multirow,booktabs}

\newenvironment{sciabstract}{%
\begin{quote} \bf}
{\end{quote}}

\newcounter{lastnote}

\title{\vspace{-60pt} Asteroseismology Can Reveal Strong Internal Magnetic Fields in Red Giant Stars}

\author
{Jim Fuller,$^{1,2\ast\dagger}$ Matteo Cantiello,$^{2\ast\dagger}$ \\
Dennis Stello,$^{3,4}$ Rafael A. Garc\'\i a,$^{5}$ Lars Bildsten$^{2,6}$\\
\\
\normalsize{$^{1}$ TAPIR, Walter Burke Institute for Theoretical Physics, Mailcode 350-17}\\ \normalsize{California Institute of Technology, Pasadena, CA 91125}\\
\normalsize{$^{2}$ Kavli Institute for Theoretical Physics, University of California, Santa Barbara, CA 93106}\\
\normalsize{$^{3}$ Sydney Institute for Astronomy (SIfA), School of Physics,}\\ \normalsize{University of Sydney, NSW 2006, Australia}\\
\normalsize{$^{4}$ Stellar Astrophysics Centre, Department of Physics and Astronomy,}\\ \normalsize{Aarhus University, Ny Munkegade 120, DK-8000 Aarhus C, Denmark}\\
\normalsize{$^{5}$ Laboratoire AIM, CEA/DSM – CNRS – Univ. Paris Diderot – IRFU/SAp}\\ \normalsize{Centre de Saclay, 91191 Gif-sur-Yvette Cedex, France}\\
\normalsize{$^{6}$ Department of Physics, University of California, Santa Barbara, CA 93106}\\
\normalsize{$^\ast$To whom correspondence should be addressed; jfuller@caltech.edu, matteo@kitp.ucsb.edu;}\\
\normalsize{$^\dagger$The first and second authors contributed equally to this work.}
}

\date{}

\begin{document} 


\baselineskip24pt


\maketitle


\begin{sciabstract}

Internal stellar magnetic fields are inaccessible to direct observations
and little is known about their amplitude, geometry and evolution. We
demonstrate that strong magnetic fields in the cores of red giant stars
can be identified with asteroseismology. The fields can
manifest themselves via depressed dipole stellar oscillation modes,
which arises from a magnetic greenhouse effect that scatters and traps
oscillation mode energy within the core of the star. The \emph{Kepler}
satellite has observed a few dozen red giants with
depressed dipole modes which we interpret as stars with
strongly magnetized cores. We find field strengths larger than
$\sim\! 10^5 \,{\rm G}$ may produce the observed
depression, and in one case we infer a minimum core field
strength of $\approx \! \! 10^7 \,{\rm G}$.

\end{sciabstract}

\section{Main Text}
\label{main}

Despite rapid progress in the discovery and characterization of magnetic fields at the surfaces of stars, very little is known about internal stellar magnetic fields. This has prevented the development of a coherent picture of stellar magnetism and the evolution of magnetic fields within stellar interiors.

\begin{figure}[h!]
\begin{center}
\includegraphics[width=0.9\columnwidth]{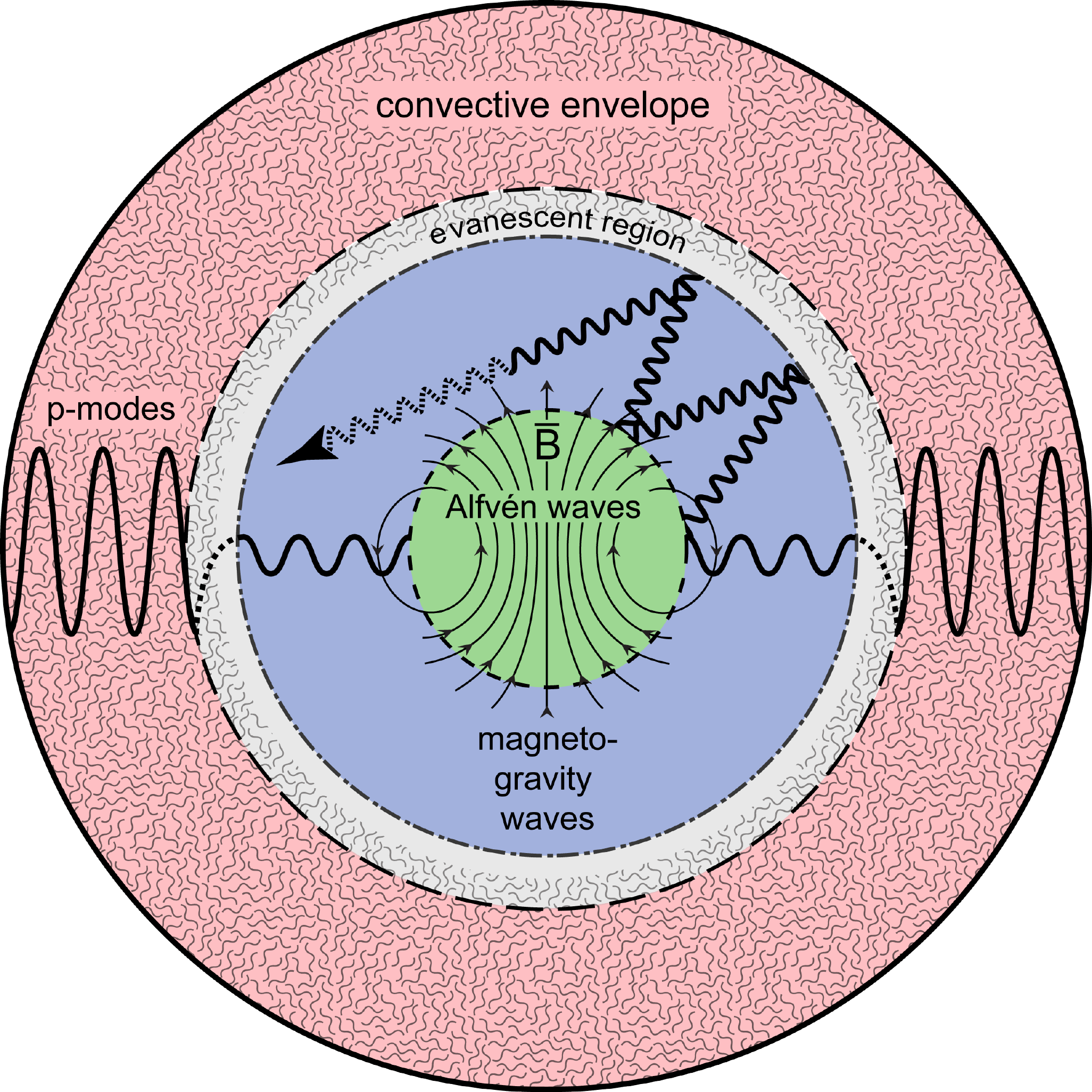}
\caption{\label{fig:cartoon} 
Wave propagation in red giants with magnetized cores. Acoustic waves excited in the envelope couple to gravity waves in the radiative core. In the presence of a magnetic field in the core, the gravity waves are scattered at regions of high field strength. Since the field cannot be spherically symmetric, the waves are scattered to high angular degree $\ell$ and become trapped within the core where they eventually dissipate (dashed wave with arrow). We refer to this as the magnetic greenhouse effect. }
\end{center}
\end{figure}

After exhausting hydrogen in their cores, most main sequence stars evolve up the red giant branch (RGB). During this phase, the stellar structure is characterized by an expanding convective envelope and a contracting radiative core. Acoustic waves (p modes) in the envelope can couple to gravity waves (g modes) in the core \cite{Bedding_2014}. Consequently, non-radial stellar oscillation modes become mixed modes that probe both the envelope (the p mode cavity) and the core (the g mode cavity), as illustrated in Fig. \ref{fig:cartoon}. Mixed modes \cite{Beck_2011} have made it possible to distinguish between hydrogen and helium-burning red giants \cite{Bedding_2011,Mosser_2014} and have been used to measure the rotation rate of red giant cores \cite{Beck_2012,Mosser_2012}.

\begin{figure}[h!]
\begin{center}
\includegraphics[width=0.9\columnwidth]{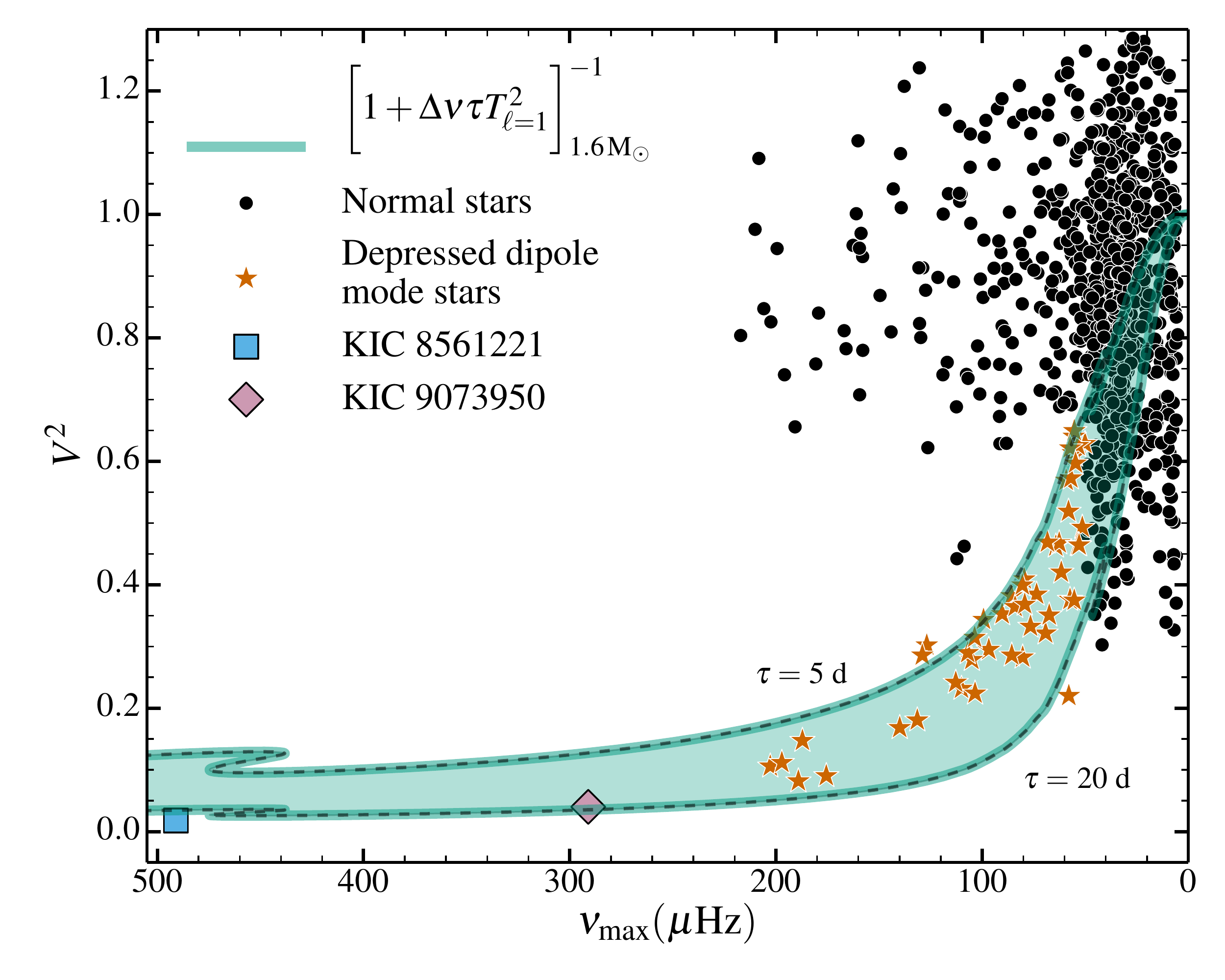}
\caption{\label{fig:moneyplot} 
Normalized visibility of dipolar modes as a function of $\nu_{\rm max}$. The star and circle symbols represent the observed visibility $V^2$ of $\ell=1$ modes \cite{supplementary}. The data is taken from \cite{Mosser_2011} and divided by 1.54 such that in normal oscillators the average visibility is $V^2=1$ \cite{Ballot_2011}. Stars with depressed dipole modes (orange stars) are identified following the same criteria as in \cite{Mosser_2011}. The theoretical band shows the visibility of depressed dipole modes in a $1.6\,M_\odot$ star, as predicted by Eqn.~\ref{eqn:vis}, and is quite insensitive to the mass of the model. The visibility of the depressed dipoles in KIC 8561221 \cite{Garcia_2014} and KIC 9073950 is also shown (square and diamond symbols, respectively). Here we used values for $\tau$  in the range 5-20 days, consistent with \cite{Dupret_2009,Corsaro_2015}.}
\end{center}
\end{figure}

A group of red giants with depressed dipole modes were identified using {\it Kepler} observations \cite{Mosser_2011}, see also Fig. \ref{fig:moneyplot}. These stars show normal radial modes (spherical harmonic degree $\ell=0$), but exhibit dipole ($\ell=1$) modes whose amplitude is much lower than usual.  Until now, the suppression mechanism was unknown \cite{Garcia_2014}. Below, we demonstrate that dipole mode suppression may result from strong magnetic fields within the cores of these red giants.

Red giant oscillation modes are standing waves that are driven by stochastic energy input from turbulent near-surface convection \cite{Goldreich_1977,Dupret_2009}. Waves excited near the stellar surface propagate downward as acoustic waves until their angular frequency $\omega$ is less than the local Lamb frequency for waves of angular degree $\ell$, i.e., until $\omega = L_{\ell} = \sqrt{\ell(\ell+1)} v_s/r$, where $v_s$ is the local sound speed and $r$ is the radial coordinate. At this boundary, part of the wave flux is reflected, and part of it tunnels into the core.

The wave resumes propagating inward as a gravity wave in the radiative core where $\omega < N$, where $N$ is the local buoyancy frequency. In normal red giants, wave energy that tunnels into the core eventually tunnels back out to produce the observed oscillation modes. We show here that suppressed modes can be explained if wave energy leaking into the core never returns back to the stellar envelope.

The degree of wave transmission between the core and envelope is determined by the tunneling integral through the intervening evanescent zone. The transmission coefficient is 
\begin{equation}
\label{eqn:integral2}
T \sim \bigg( \frac{r_1}{r_2} \bigg)^{\sqrt{\ell(\ell+1)}} \, ,
\end{equation}
where $r_1$ and $r_2$ are the lower and upper boundaries of the evanescent zone, respectively. The fraction of wave energy transmitted through the evanescent zone is $T^2$. For waves of the same frequency, larger values of $\ell$ have larger values of $r_2$, thus Eqn. \ref{eqn:integral2} demonstrates that high $\ell$ waves have much smaller transmission coefficients through the evanescent zone. 

The visibility of stellar oscillations depends on the interplay between driving and damping of the modes \cite{Dupret_2009,Benomar_2014}. To estimate the reduced mode visibility due to energy loss in the core, we assume that all mode energy which leaks into the g mode cavity is completely lost. The mode then loses a fraction $T^2$ of its energy in a time $2 t_{\rm cross}$, where $t_{\rm cross}$ is the wave crossing time of the acoustic cavity. Due to the larger energy loss rate, the mode has less energy $E_{\rm ac}$ within the acoustic cavity and produces a smaller luminosity fluctuation $V$ at the stellar surface, whose amplitude scales as $V^2 \propto E_{\rm ac}$. We show \cite{supplementary} that the ratio of visibility between a suppressed mode $V_{\rm sup}$ and its normal counterpart $V_{\rm norm}$ is 
\begin{equation}
\label{eqn:vis}
\frac{V_{\rm sup}^2}{V_{\rm norm}^2} = \bigg[1 + \Delta \nu \, \tau \, T^2 \bigg]^{-1} \, ,
\end{equation}
where $\Delta \nu \simeq (2 t_{\rm cross})^{-1}$ \cite{Chaplin_2013} is the large frequency separation between acoustic overtone modes, and $\tau$ is the damping time of a radial mode with similar frequency. We evaluate $T^2$ from our stellar models using Eqn. \ref{eqn:integral}, whereas $\tau \approx 5 \! - \! 10 \, {\rm days}$  \cite{Dupret_2009,Corsaro_2012,Grosjean_2014,Corsaro_2015} for stars ascending the RGB.

Most observed modes  are near the frequency  $\nu_{\rm max}$, which is determined by the evolutionary state of the star. On the RGB, more evolved stars generally have smaller $\nu_{\rm max}$. Fig. \ref{fig:moneyplot} compares our estimate for suppressed dipole mode visibility (Eqn. \ref{eqn:vis}) with {\it Kepler} observations \cite{Mosser_2011,Garcia_2014}. The objects identified by \cite{Mosser_2011} as depressed dipole mode stars lie very close to our estimate. The striking agreement holds over a large baseline in $\nu_{\rm max}$ extending from the very early red giants KIC8561221 \cite{Garcia_2014} and KIC9073950 at high $\nu_{\rm max}$ to near the luminosity bump at low $\nu_{\rm max}$. The observations are consistent with nearly total wave energy loss in the core, as partial energy loss would create stars with less depressed modes, which seem to be rare.

We conclude that the cores of stars with depressed dipole modes efficiently trap or disrupt waves tunneling through the evanescent region. This is further supported by their normal $\ell=0$ mode visibility, because radial modes do not propagate within the inner core and because much larger field strengths are required to alter acoustic waves. The absence (or perhaps smaller degree) of  depression observed for $\ell=2$ modes \cite{Mosser_2011} occurs because quadrupole modes have a smaller transmission coefficient $T$, and less of their energy leaks into the core.

An additional consequence is that the larger effective damping rate for depressed modes will lead to larger line widths in the oscillation power spectrum. The linewidth of a depressed dipole mode is $\tau^{-1} + \Delta \nu T^2_{\ell}$ and is generally much larger than that of a normal mode. The depressed dipole modes in KIC8561221 \cite{Garcia_2014} indeed have much larger linewidths than normal dipole modes in similar stars.

Magnetic fields can provide the mechanism for trapping oscillation mode energy in the core by altering gravity wave propagation. The nearly horizontal motions and short radial wavelengths of gravity waves in RGB cores will bend radial magnetic field lines, creating strong magnetic tension forces. The acceleration required to restore a wave of angular frequency $\omega$ and horizontal displacement $\xi_{\rm h}$ is $\xi_{\rm h} \omega^2$, whereas the magnetic tension acceleration due to a radial magnetic field of strength $B_r$ is $\xi_{\rm h} B_r^2 k_r^2/(4 \pi \rho)$, where $k_r$ is the radial wavenumber and $\rho$ is the density. Gravity waves are strongly altered by the magnetic fields when the magnetic tension force dominates, which for dipole waves occurs at a critical magnetic field strength \cite{supplementary}
\begin{equation}\label{eqn:Bc}
B_c= \sqrt{\frac{\pi \rho}{2}} \, \frac{\omega^2 r}{N} \, .
\end{equation}
This field strength approximately corresponds to the point at which the Alfv\'en speed becomes larger than the radial group velocity of gravity waves.

\begin{figure}[h!]
\begin{center}
\includegraphics[width=0.75\columnwidth]{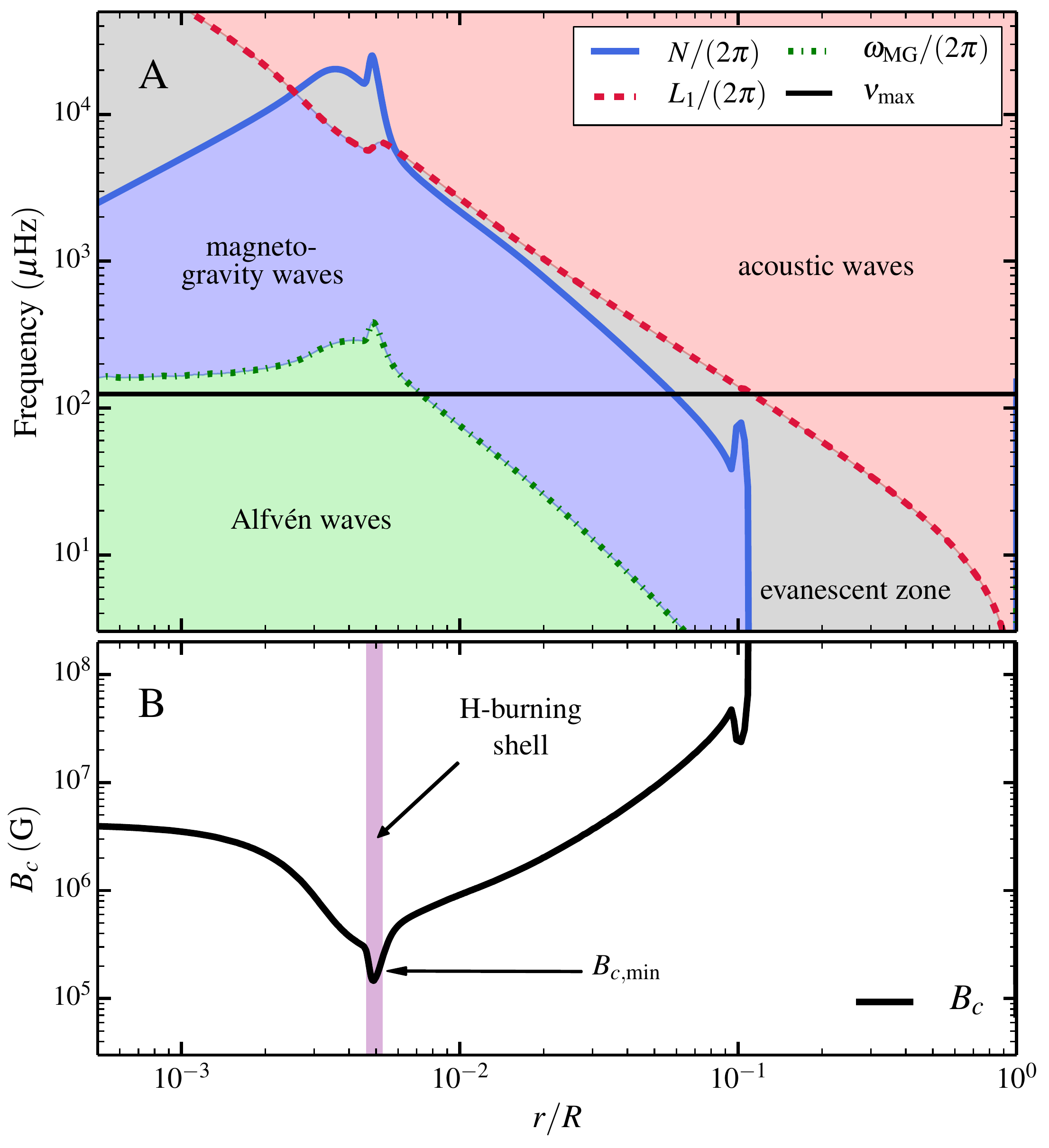}
\caption{\label{fig:Prop}
Propagation diagram for a magnetized red giant model. The model has $M = 1.6 \, M_\odot$, $R =6.6 \, R_\odot$, ${\nu}_{\rm max} = 120 \, {\rm {\mathrm{\mu}}Hz}$, and a core magnetic field of $\approx 6 \times 10^6 \,{\rm G}$ (see Fig. \ref{Fig:Struc}). ({\bf A}) The red, blue, and green lines are the dipole Lamb frequency $L_1$, the buoyancy frequency $N$, and the magneto-gravity frequency $\omega_{\rm MG}$ (defined in Eqn. \ref{eqn:maggrav}), respectively. Regions are colored by the types of waves they support: the red region is the acoustic wave cavity, the blue region is the magneto-gravity wave cavity, and the green region hosts only Alfv\'en waves. The horizontal line is the frequency of maximum power, $\nu_{\rm max}$, for this stellar model. Waves at this frequency behave like acoustic waves near the surface, magneto-gravity waves in the outer core, and Alfv\'en waves in the inner core. ({\bf B}) Critical radial magnetic field strength $B_c$ needed to suppress dipole modes. $B_c$ (Eqn. \ref{eqn:Bc}) is evaluated at the angular frequency $\omega \! = \! 2 \pi \nu_{\rm max}$. $B_c$ has a sharp minimum at the H-burning shell, which determines the minimum field strength $B_{c,{\rm min}}$ required for dipole mode suppression.}
\end{center}
\end{figure}

Magneto-gravity waves cannot exist in regions with $B_r \! > \! B_c$ where magnetic tension overwhelms the buoyancy force, i.e., the stiff field lines cannot be bent by the placid gravity wave motion. Consequently, dipole magneto-gravity waves become evanescent when $\omega \! < \! \omega_{\rm MG}$,  where the magneto-gravity frequency $\omega_{\rm MG}$ is defined as
\begin{equation}
\label{eqn:maggrav}
\omega_{\rm MG} = \bigg[ \frac{2}{\pi} \frac{B_r^2 N^2}{\rho r^2} \bigg]^{1/4} \, .
\end{equation}
Fig. \ref{fig:Prop} shows a wave propagation diagram in which a strong internal magnetic field prevents magneto-gravity wave propagation in the core.

In red giants, $B_c$ is typically smallest at the peak in $N$ corresponding to the sharp density gradient within the hydrogen burning (H-burning) shell. Therefore, gravity waves are most susceptible to magnetic alteration in the H-burning shell, and the observation of a star with depressed dipole modes thus provides a {\it lower limit} to the radial field strength (Eqn. \ref{eqn:Bc}) evaluated in the H-burning shell. We refer to this field strength as $B_{c,{\rm min}}$. Magnetic suppression via horizontal fields can also occur, but in general requires much larger field strengths.
  
In stars with field strengths exceeding $B_c$ (Eqn. \ref{eqn:Bc}) somewhere in their core, incoming dipole gravity waves will become evanescent where $B_r \! > \! B_c$. At this point, the waves must either reflect or be transmitted into the strongly magnetized region as Alfv\'en waves. In either case, the reflection/transmission process modifies the angular structure of the waves such that their energy is spread over a broad spectrum of $\ell$ values \cite{supplementary}. Once a dipole wave has its energy transferred to higher values of $\ell$, it will not substantially contribute to observable oscillations at the stellar surface, because higher $\ell$ waves are trapped within the radiative core by a thicker evanescent region (see Eqn. \ref{eqn:integral2} and Fig. \ref{fig:cartoon}) separating the core from the envelope. Even if some wave energy does eventually return to the surface to create an oscillation mode, the increased time spent in the core results in a very large mode inertia, greatly reducing the mode visibility. Additionally, high $\ell$ waves will not be detected in {\it Kepler} data due to the geometric cancellation which makes $\ell \! \gtrsim \! 3$ modes nearly invisible \cite{Bedding_2010}.

The magnetic greenhouse effect arises not from the alteration of incoming wave frequencies, but rather due to modification of the wave angular structure. Such angular modification originates from the inherently non-spherical structure (because $\nabla \cdot {\bf B} = 0$) of even the simplest magnetic field configurations.

Dipole oscillation modes can be suppressed if the magnetic field strength exceeds $B_c$ (Eqn. \ref{eqn:Bc}) at some point within the core. We therefore posit that stars with depressed dipole oscillation modes have minimum core field strengths of $B_{c,{\rm min}}$. Stars with normal dipole oscillation modes cannot have radial field strengths in excess of $B_{c,{\rm min}}$ within their H-burning shells. However, they may contain larger fields away from the H-burning shell, or they may contain fields that are primarily horizontal (e.g., strong toroidal fields).

\begin{figure}[h!]
\begin{center}
\includegraphics[width=0.75\columnwidth]{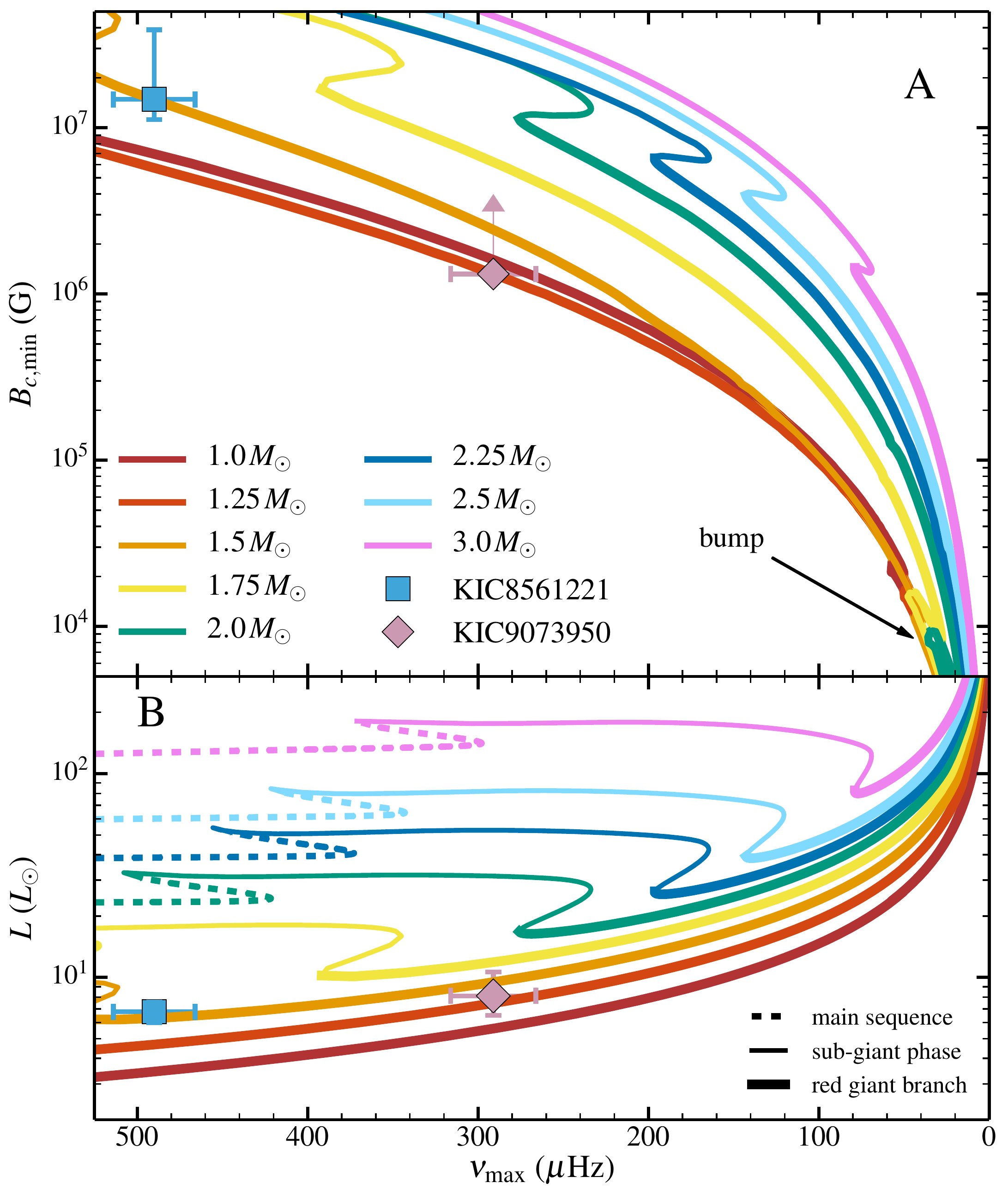}
\caption{\label{fig:Bc}
Minimum field strength $B_{c,{\rm min}}$ required for magnetic suppression of dipole oscillation modes in stars evolving up the RGB. ({\bf A}) $B_{c,{\rm min}}$ is shown as a function of the frequency of maximum power, $\nu_{\rm max}$, for stars of different mass. $B_{c,{\rm min}}$ has been computed for modes with angular frequency $\omega \! = \! 2 \pi \nu_{\rm max}$. We have labeled the observed value of $\nu_{\rm max}$ and inferred core field strength $B_r$ for the lower RGB star KIC8561221 \cite{Garcia_2014} and a lower limit to the field strength in KIC9073950. ({\bf B}) Asteroseismic HR diagram showing stellar evolution tracks over the same range in $\nu_{\rm max}$. Stars evolve from left to right as they ascend the RGB. Dashed lines show the end of main sequence evolution, thin solid lines are the sub-giant phase, and thick solid lines are the RGB. We do not show any evolution beyond the RGB (i.e., no clump or asymptotic giant branch).}
\end{center}
\end{figure}

Fig. \ref{fig:Bc} shows the value of $B_{c,{\rm min}}$ as stars evolve up the RGB. We have calculated $B_{c,{\rm min}}$ for angular frequencies $\omega \! = \! 2 \pi \nu_{\rm max}$, and evaluated $\nu_{\rm max}$ using the scaling relation proposed by \cite{Brown_1991}. On the lower RGB, where $\nu_{\rm max} \! \gtrsim \! 250\,\mu$Hz, field strengths of order $B_{c,{\rm min}} \! \gtrsim 10^6 \, {\rm G}$ are required for magnetic suppression. As stars evolve up the red giant branch, the value of $B_{c,{\rm min}}$ decreases sharply, primarily because $\nu_{\rm max}$ decreases. By the luminosity bump (near $\nu_{\rm max} \! \sim 40\,\mu$Hz), field strengths of only $B_{c,{\rm min}} \! \sim \! \!10^4 \, {\rm G}$ are sufficient for magnetic suppression. Magnetic suppression during the sub-giant phase (higher $\nu_{\rm max}$) and in higher mass stars ($M \! \gtrsim \! 2 M_\odot$) may be less common due to the larger field strengths required.

For a given field strength, there is a transition frequency $\nu_c$ below which modes will be strongly suppressed and above which modes will appear normal. Stars which show this transition are especially useful because they allow for an inference of $B_r$ at the H-burning shell via Eqn. \ref{eqn:Bc}, evaluated at the transition frequency $\omega \! = \! 2 \pi \nu_c$.  The RGB star KIC8561221 shows this transition \cite{Garcia_2014}. Using the observed value of $\nu_c \! \approx \! 600 \, \mu{\rm Hz}$, we infer that the radial component of the magnetic field within the H-burning shell is $B_r \! \approx \! 1.5 \times 10^{7} \, {\rm G}$, although we cannot rule out the presence of stronger fields away from the H-burning shell. This large field strength may indicate KIC8561221 is the descendant of a magnetic Ap star whose internal field was much stronger than the typical surface fields of $B \! \sim \! 3 \, {\rm kG}$ of Ap stars \cite{Auri_re_2007}.

In principle, it is possible that another symmetry-breaking mechanism within the core could suppress dipole mode amplitudes. The only other plausible candidate is rapid core rotation. In order for rotation to strongly modify the incoming waves such that they will be trapped in the core, the core must rotate at a frequency comparable to $\nu_{\rm max}$, roughly two orders of magnitude faster than the values commonly measured in red giant cores \cite{Beck_2011,Mosser_2012,deheuvels_2014}. The depressed dipole mode star KIC8561221 \cite{Garcia_2014} does not exhibit rapid core rotation and disfavors the rotation scenario.

A magnetic field of amplitude $B \! > \! 10^4 \, {\rm G}$ (see Fig. \ref{fig:Bc}) could be present in the core of a red giant if it was retained from previous phases of stellar formation/evolution \cite{supplementary}. These strong fields may reside within the inner core with little external manifestation apart from the reduced visibility of the dipole modes. However, fields of similar amplitude have been discussed in order to explain the suppression of thermohaline mixing in a small fraction of red giant stars, as inferred from the observations of their surface abundances \cite{Charbonnel_2007}. The inferred core field strength of $B_r \! \gtrsim \! 1.5 \! \times \! 10^7 \, {\rm G}$ in KIC8561221 shows very strong magnetic fields ($B \! \gg \! 10^6 \, {\rm G}$) can exist within the radiative cores of early RGB stars. Since these fields are likely inherited from previous stages of stellar evolution, slightly weaker ($B \! \gg \! 10^5 \,{\rm G})$ fields could exist in the cores of exceptional very highly magnetized main sequence stars.

\bibliography{refs}

%\begin{scilastnote}
All the authors of this paper thank the KITP and the organizers of the ``Galactic Archaeology and Precision Stellar Astrophysics{\textquotedblright} program held from January to April 2015. JF acknowledges partial support from NSF under grant no. AST-1205732 and through a Lee DuBridge Fellowship at Caltech. RAG acknowledge the support of the European Community{'}s Seventh Framework Programme (FP7/2007-2013) under grant agreement No. 312844 (SPACEINN), and from the CNES. This project was supported by NASA under TCAN grant number NNX14AB53G and the NSF under grants PHY 11-25915 and AST 11-09174. %\end{scilastnote}
  
\ \\
{\bf Supplementary Materials} \\
\ \\
Supplementary Text \\
Figs. S1 to S3 \\
References (23-56) \\

\newpage

\setcounter{equation}{0}
\setcounter{figure}{0}
\setcounter{table}{0}
\setcounter{section}{0}
\setcounter{page}{1}
\renewcommand{\theequation}{S\arabic{equation}}
\renewcommand{\thefigure}{S\arabic{figure}}
\renewcommand{\thetable}{S\arabic{table}}
\renewcommand{\thesection}{S\arabic{section}}
\renewcommand{\thepage}{\arabic{page}}

\begin{center}
\includegraphics[width=1.0\columnwidth]{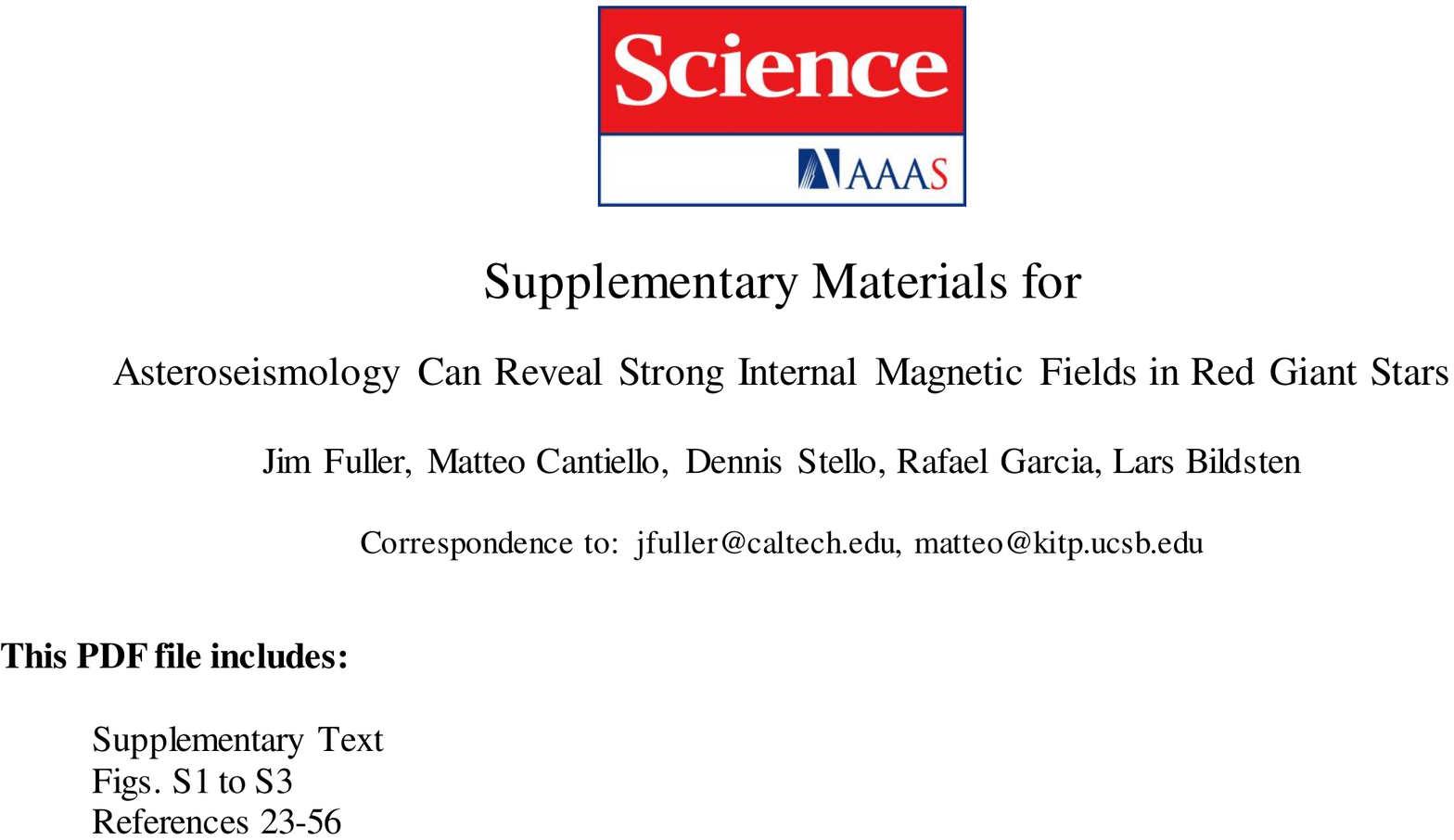}
\end{center}

\newpage

\section{Supplementary Text}

\subsection{Stellar Models}

We have used the Modules for Experiments in Stellar Evolution (MESA, release 7385) code \cite{Paxton_2010,Paxton_2013} to evolve low-mass stars with initial mass in the range 1-3.0$\,M_\odot$. Models have been evolved from the pre-main-sequence to the tip of the red giant branch. We chose a metallicity of $Z=0.02$ with a mixture taken from  \cite{2005ASPC..336...25A}; the plasma opacity is determined  using the OPAL opacity tables from \cite{Iglesias_1996}. Convective regions are calculated using the mixing-length theory (MLT) with $\alpha_{\rm MLT} = 2.0$. The boundaries of convective regions are determined using the Ledoux criterion. Overshooting is parameterized by an exponentially decaying diffusivity that decays over a distance $f_{\rm ov} H$ above the convective boundary \cite{2000A&A...360..952H}, with $f_{\rm ov}= 0.018$. We include in Section \ref{inlist} the inlist used for running the calculations.

To estimate plausible magnetic field strengths within the cores of red giants, we consider two scenarios: one based on observational evidence of surface {fi}elds, and one based on MHD modeling of magnetic {fi}elds in convective stellar cores. First, we extrapolate inward from a main sequence surface field of $B \! \sim \! 3\, {\rm kG}$, as appropriate for magnetic Ap stars \cite{Auri_re_2007}, assuming the field is a pure dipole such that the field strength scales as $B \! \propto \! r^{-3}$. Since the radius of the convective core is typically $r_c \! \sim \! R/10$ for low mass main sequence stars, field strengths of $B \! > \! 10^{5} \, {\rm G}$ are attainable near the core. 

Second, we estimate field strengths produced by a magnetic dynamo that operates within the convective core of a star while it is on the main sequence. In this case, MHD simulations suggest equipartition (and even super-equipartition) magnetic field strengths may be generated \cite{Featherstone_2009}, i.e., magnetic fields whose energy density is comparable to that of the kinetic energy of convective flows such that
\begin{equation}
\label{eqn:beq}
\frac{B^2}{8 \pi} \sim \rho v_{\rm con}^2 \, .
\end{equation}
We calculate typical core convective velocities $v_{\rm con}$ using mixing length theory. In our stellar models, we evaluate Eqn. \ref{eqn:beq} to find that core magnetic fields of $B \! \sim \! 3 \! \times \! 10^5 \, {\rm G}$ could be generated during the main sequence.

To extrapolate to field strengths plausibly obtained within the radiative cores of red giants, we assume that the magnetic flux (calculated via the methods above) within the core is conserved as it contracts. This is a good approximation for stable magnetic equilibria discussed here because the timescale for the field to diffuse through the star (the Ohmic timescale) is longer than the main sequence timescale. At each mass shell within a red giant, the field strength is then approximated by
\begin{equation}
\label{eqn:BRG}
B_{\rm RG} = \bigg(\frac{r_{\rm MS}}{r_{\rm RG}}\bigg)^2 B_{\rm MS} \, ,
\end{equation}
where $B_{\rm RG}$ is the field strength while on the RGB, $r_{\rm RG}$ and $r_{\rm MS}$ are the radial coordinates of the shell on the RGB and MS, respectively, and $B_{\rm MS}$ is the MS field strength. Mass shells enclosing $M \sim 0.2 \, M_\odot$ (which are located just outside the MS core and near the H-burning shell on the lower RGB) typically contract by a factor of a few from the MS to the lower RGB. The magnetic field may therefore be amplified and field strengths in excess of $10^6 \, {\rm G}$ are quite plausible within the H-burning shells of RGB stars. 

\begin{figure}[h!]
\begin{center}
\includegraphics[width=0.85\columnwidth]{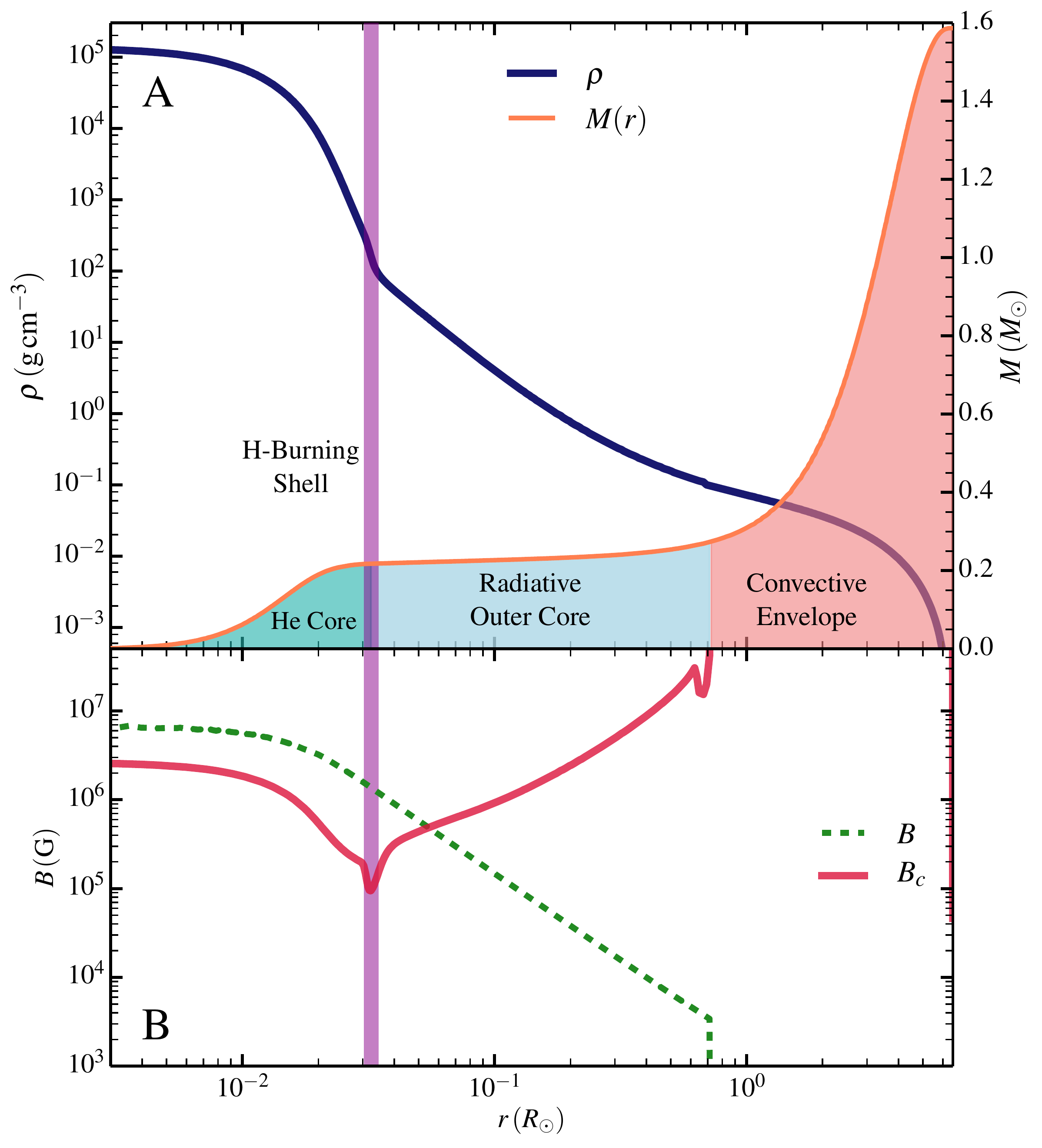}
\caption{\label{Fig:Struc}
Structure of the $M \! = \! 1.6 \, M_\odot$ stellar model shown in Fig. \ref{fig:Prop}. ({\bf A}) At this stage of evolution, the star has a $\sim \! 0.2 \, M_\odot$ helium core, surrounded by a $\sim \! 0.1 \, M_\odot$ radiative outer core. The vertical purple band shows the location of the hydrogen-burning shell. The bulk of the mass and radial extent of the star is comprised by the thick convective envelope. ({\bf B}) Core magnetic field. The field strength $B$ is calculated by assuming magnetic flux conservation from a dipole field on the main sequence as described in the supplementary material. We also plot the critical field $B_c$ from Eqn. \ref{eqn:Bc} as in Fig. \ref{fig:Prop}. Because $B \! > \! B_c$ in the core of this stellar model, the magnetic greenhouse effect may occur.}
\end{center}
\end{figure}

Fig. \ref{Fig:Struc} shows the density, mass, and magnetic field profiles of the $1.6 \, M_\odot$ stellar model used to generate Fig. \ref{fig:Prop}. To make this model, we extrapolate a dipole field inward from a surface value of $3 \! \times \! 10^{3} \, {\rm G}$ (as described above), with an artificial cap at a field strength of $5 \! \times \! 10^{4} \, {\rm G}$. We then calculate the corresponding RGB field profile using the flux conservation described above (for simplicity we set the field equal to zero in convective regions of the RGB model). This relatively conservative approach yields a field strength of $\sim \! \! 10^6 \, {\rm G}$ at the H-burning shell, sufficient for magnetic suppression of dipole oscillation modes. We note that field strengths of this magnitude are orders of magnitude below equipartition with the gas pressure, and therefore have a negligible influence on the stellar structure.

\subsection{Mode Visibility}

Here we estimate the visibility of modes suppressed via the magnetic greenhouse effect. To do this, we consider the energy balance between driving and damping of a mode. Each mode receives a stochastic energy input $\dot{E}_{\rm in}$ \cite{Dupret_2009}. At its time-averaged equilibrium amplitude, the energy input and damping rates of the mode are equal, such that
\begin{equation}
\label{eqn:enorm}
\dot{E}_{\rm in} = \dot{E}_{\rm out} = E_{\alpha} \gamma_\alpha \, ,
\end{equation}
where $E_{\alpha}$ is the energy contained in the mode and $\gamma_\alpha$ is its damping rate. 

Modes suppressed via the magnetic greenhouse effect have an extra source of damping determined by the rate at which energy leaks through the evanescent region separating the acoustic cavity from the g mode cavity. For suppressed modes, we assume that any mode energy which leaks into the g mode cavity is completely lost via the magnetic greenhouse effect. Similar calculations have been employed for more evolved red giants where waves entering the core are damped by radiative diffusion \cite{Dziembowski_1977,Osaki_1977,Dziembowski_2012}. Given some mode energy contained within the acoustic cavity, $E_{\rm ac}$, the rate at which mode energy leaks into the core is
\begin{equation}
\dot{E}_{\rm leak} = \frac{E_{\rm ac}}{t_{\rm leak}} ,
\end{equation}
where $t_{\rm leak}$ is the time scale on which mode energy leaks into the core. As explained in the text, the energy leakage time scale is
\begin{equation}
\label{eqn:tleak}
t_{\rm leak} = \frac{2 t_{\rm cross}}{T^2} \, ,
\end{equation}
where the transmission coefficient $T$ is
\begin{equation}
\label{eqn:integral}
T =  \exp{ \bigg[ \int^{r_2}_{r_1} i k_r dr} \bigg] \, .
\end{equation}
The value of $T$ is approximately the fractional decrease in wave amplitude across the evanescent region, whereas $T^2$ is the fractional decrease in wave energy. In the WKB approximation, the value of the radial wavenumber $k_r$ within the evanescent region is
\begin{equation}
\label{eqn:kr}
k_r^2 = \frac{ \big(L_\ell^2 - \omega^2 \big) \big(N^2 - \omega^2 \big) }{v_s^2 \omega^2} \, .
\end{equation}
Deep within an evanescent region where $N^2 \! \ll \! \omega \! \ll \! L_\ell^2$, Eqn. \ref{eqn:kr} evaluates to $k_r \! \sim \! i \sqrt{\ell (\ell +1)}/r$, and Eqn. \ref{eqn:integral} yields the expression for $T$ in Eqn. \ref{eqn:integral2}. The wave crossing time for acoustic waves is
\begin{equation}
t_{\rm cross} = \int^R_{r_2} \frac{dr}{v_s} \, .
\end{equation}

A suppressed mode is also damped by the same mechanisms as a normal mode. In the case of envelope modes for stars low on the RGB, this damping is created by convective motions near the surface of the star \cite{Houdek_1999,Dupret_2009}. The equilibrium energy of the suppressed mode is
\begin{equation}
\label{eqn:esup}
\dot{E}_{\rm in} = \dot{E}_{\rm out} = E_{\rm ac} \bigg[\gamma_{\rm ac} +  \frac{T^2}{2 t_{\rm cross}} \bigg],
\end{equation}
where $\gamma_{\rm ac}$ is the damping rate due to convective motions in the acoustic cavity.

Now, we assume that the suppression mechanism is localized to the core and that the energy input $\dot{E}_{\rm in}$ is unaltered. Then we can set Eqns. \ref{eqn:enorm} and \ref{eqn:esup} equal to each other to find
\begin{equation}
\label{edamp}
E_{\alpha} \gamma_\alpha =  E_{\rm ac} \bigg[\gamma_{\rm ac} +  \frac{T^2}{2 t_{\rm cross}} \bigg] \, .
\end{equation}
The damping of a normal mode is localized to the acoustic cavity, so its energy loss rate can be written
\begin{equation}
{E_\alpha} \gamma_\alpha \simeq E_{\alpha,{\rm ac}} \gamma_{\rm ac} \, ,
\end{equation}
where $E_{\alpha,{\rm ac}}$ is the mode energy contained in the acoustic cavity. Inserting this into Eqn. \ref{edamp}, we have
\begin{equation}
\label{eqn:ebalance}
E_{\alpha,{\rm ac}} \gamma_{\rm ac} =  E_{\rm ac} \bigg[\gamma_{\rm ac} +  \frac{T^2}{2 t_{\rm cross}} \bigg] \, .
\end{equation}

The energy of a mode within the envelope is proportional to its surface amplitude squared, hence, the visibility of a mode scales as $V_{\alpha}^2 \propto E_{\alpha,{\rm ac}}$. Then the ratio of the visibility of the suppressed mode to that of the normal mode is
\begin{equation}
\frac{V_{\rm sup}^2}{V_{\rm norm}^2} = \frac{E_{\rm ac}}{E_{\alpha,{\rm ac}}} \, .
\end{equation}
Then Eqn. \ref{eqn:ebalance} leads to
\begin{equation}
\frac{V_{\rm sup}^2}{V_{\rm norm}^2} = \frac{\gamma_{\rm ac}}{\gamma_{\rm ac} + T^2/(2 t_{\rm cross})} \, .
\end{equation}
Using the fact that the large frequency separation is $\Delta \nu \simeq (2 t_{\rm cross})^{-1}$ \cite{Chaplin_2013} and defining $\tau = \gamma_{\rm ac}^{-1}$, we have our final result:
\begin{equation}
\frac{V_{\rm sup}^2}{V_{\rm norm}^2} = \bigg[1 + \Delta \nu \tau T^2 \bigg]^{-1} \, .
\end{equation}
The damping time $\tau$ is the lifetime of wave energy located in the acoustic cavity. It is not equal to the lifetime of a normal dipole mode, because much of the dipole mode energy resides within the core. Instead, $\tau$ is approximately equal to the lifetime of a radial mode, because all of its energy is in the acoustic cavity. Thus, $\tau$ can be equated with observed/theoretical lifetimes of radial modes.
 
We emphasize that the magnetic greenhouse mechanism that operates in depressed dipole oscillators is not directly related to the suppression of solar-like oscillations in stars exhibiting surface magnetic activity \cite{Garcia_2010,Chaplin_2011,Gaulme_2014}. In the latter case, the depression likely arises from magnetic effects in the convective envelope quenching the amplitudes of all oscillation modes, not just the dipole modes.

\subsection{Wave Leakage Time}

The wave leakage time scale $t_{\rm leak}$ on which wave energy tunnels from the acoustic cavity into the stellar core can be estimated from Eqn. \ref{eqn:tleak}, with the value of $T$ calculated from the first part of Eqn. \ref{eqn:integral}, or approximated from Eqn. \ref{eqn:integral2}.

To compute a more precise estimate, we solve the forced adiabatic linearized hydrodynamic (non-magnetic) wave equations for our stellar models (using the Cowling approximation), assuming all wave energy that tunnels into the outer core is lost within the inner core. To do this, we place the inner boundary of our computational grid at a radius $r/R = 0.01$, which is always within the stably stratified regions of our red giant models. We then impose a radiative inner boundary condition \cite{Fuller_2012}. On the outer boundary, we impose a forcing/normalization condition on the real part of the wave displacement vector {\boldmath $\xi$}, i.e., we set ${\rm Re} \big( \xi_r \big) =1$, where $\xi_r$ is the radial component of the displacement vector. For the imaginary component, we adopt the standard reflective outer boundary condition, ${\rm Im} \big( \delta P \big) = \rho g {\rm Im} \big( \xi_r \big)$, where $\delta P$ is the Eulerian pressure perturbation, $\rho$ is the density, and $g$ is the gravitational acceleration. Physically, this scenario represents the forcing of the stellar surface at a given angular frequency $\omega$, and the eventual leakage of the wave energy into the core of the star.

After solving the wave equations, we compute the leakage time of the wave energy contained within the acoustic cavity at radii $r_2 \! < r \! < R$. The wave energy contained within the acoustic cavity is simply
\begin{equation}
\label{eqn:ecav}
E_{\rm ac} = \int^{R}_{r_2} dr \, \rho r^2 \omega^2 \bigg( | \xi_r |^2 + \ell(\ell+1) | \xi_\perp |^2 \bigg) \, ,
\end{equation}
where $\xi_\perp$ is the horizontal component of the wave displacement vector. The rate at which energy leaks through the inner boundary is
\begin{equation}
\label{eqn:eleak}
\dot{E}_{\rm leak} = \rho r^3 \omega^3 \bigg[ {\rm Re}\big( \xi_\perp \big) {\rm Im}\big( \xi_r \big) - {\rm Re}\big( \xi_r \big) {\rm Im} \big( \xi_\perp \big) \bigg] \, ,
\end{equation}
evaluated at the inner boundary of the grid. The wave energy leakage time is then
\begin{equation}
\label{eqn:tleak2}
t_{\rm leak} = \frac{ E_{\rm ac} }{\dot{E}_{\rm leak}} \, .
\end{equation}
We have calculated the leakage timescales for waves with frequencies near $\omega_{\rm max} \! = \! 2 \pi \nu_{\rm max}$ for stellar models on the RGB. For this computational technique, the energy $E_{\rm ac}$ contained within the acoustic cavity peaks at the p mode frequencies of the stellar model. The energy loss rate $\dot{E}_{\rm leak}$ also peaks at the mode frequencies, so that the value of $t_{\rm leak}$ is essentially independent of $\omega$.

\begin{figure}[h!]
\begin{center}
\includegraphics[width=0.9\columnwidth]{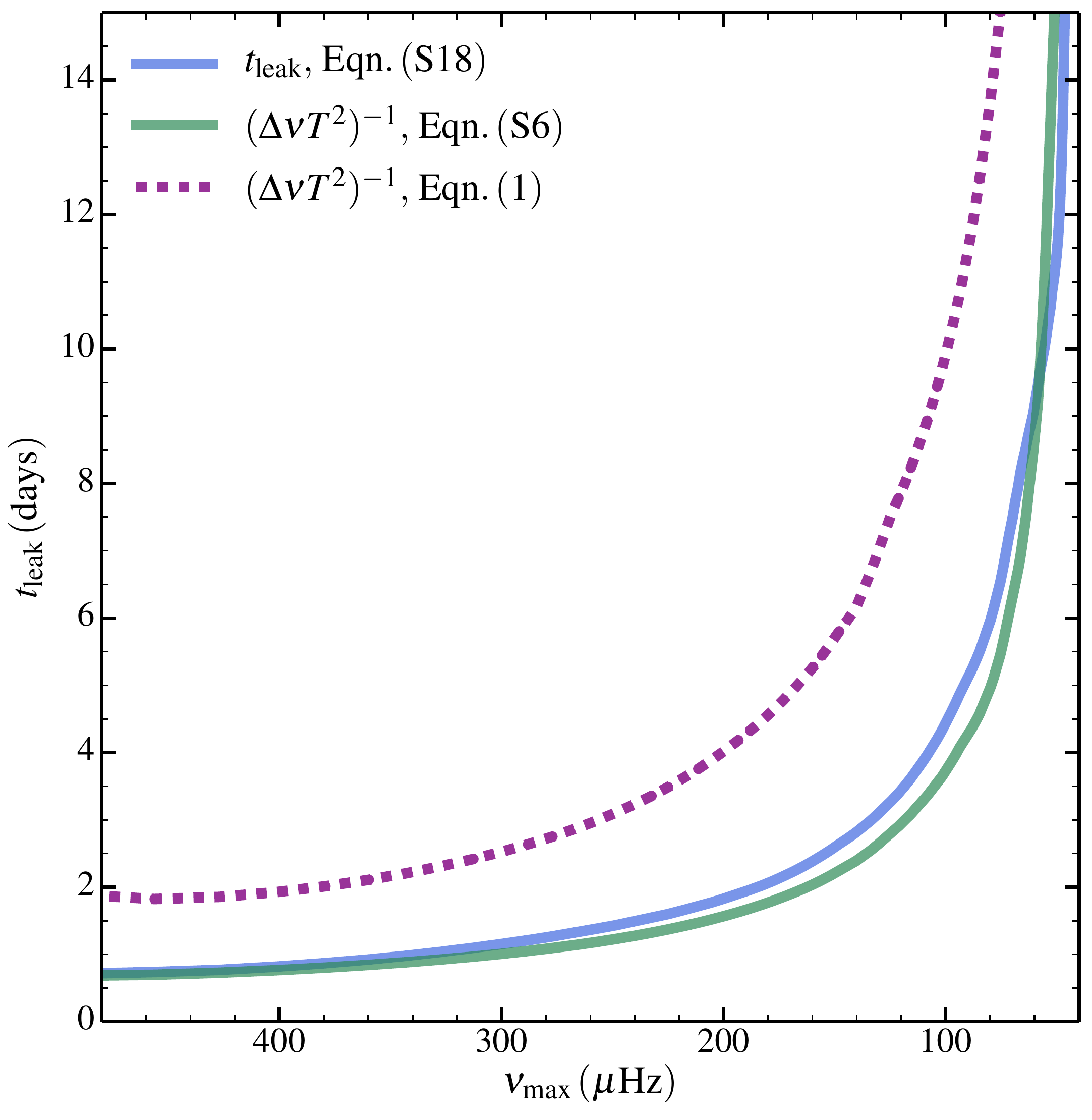}
\caption{\label{fig:DipoleTime}
Wave leakage time scale as a function of $\nu_{\rm max}$ for a $1.6 \, M_\odot$ stellar model. The leakage timescale has been calculated by solving the linearized wave equations (Eqn. \ref{eqn:tleak2}), and by approximating it using Eqn. \ref{eqn:tleak}, evaluating the transmission coefficient $T$ via Eqn. \ref{eqn:integral2} or \ref{eqn:integral}.}
\end{center}
\end{figure}

Fig. \ref{fig:DipoleTime} shows the exact value of $t_{\rm leak}$ calculated from Eqn. \ref{eqn:tleak2}, and $t_{\rm leak}$ approximated from Eqn. \ref{eqn:tleak}, with $T$ calculated via Eqn.s \ref{eqn:integral2} and \ref{eqn:integral}. Clearly, evaluating $t_{\rm leak}$ via Eqn. \ref{eqn:tleak} with $T$ calculated from Eqn. \ref{eqn:integral} is a very good approximation, accurate to within $\sim \! 10 \%$ for our stellar models. However, using the approximation of Eqn. \ref{eqn:integral2} is not very accurate, and generally produces a value of $t_{\rm leak}$ too large by a factor of $\sim \! 2$. We conclude that we may accurately estimate mode visibilities using Eqn. \ref{eqn:vis}, so long as the value of $T$ is calculated with an integral over the evanescent region as in Eqn. \ref{eqn:integral}. The approximation of $T$ in Eqn. \ref{eqn:integral2} should not be used for visibility calculations, although it is still useful because it demonstrates the scaling of $T$ with wave angular degree $\ell$ and the size of the evanescent region.

\subsection{Magneto-Gravity Waves}

The properties of magnetohydrodynamic waves in red giant cores can be understood using a local (WKB) analysis for high wavenumbers ${\bf k}$, in which $k r \gg 1$ and $k H \gg 1$ (where $H$ is a pressure scale height). We show below the WKB limit is a good approximation in our stellar models. In what follows, we shall also use the adiabatic, anelastic, and ideal MHD approximations, which are all valid for the magneto-gravity waves we consider in red giant cores.

Using the approximations above, the dispersion relation for MHD waves is \cite{unno:89}
\begin{equation}
\label{eqn:disp}
\bigg( \omega^2 - \omega_A^2 \bigg) \bigg( \omega^2 - \frac{k_\perp^2}{k^2}N^2 - \omega_A^2 \bigg) = 0.
\end{equation}
Here, $\omega$ is the angular frequency of the wave, $k_\perp = \sqrt{l(l+1)}/r$ is the horizontal wavenumber, $N$ is the Brunt-V\"{a}is\"{a}l\"{a} (buoyancy) frequency, and the Alfv\'en frequency is
\begin{equation}
\label{eqn:alven}
\omega_A^2 = \frac{ \big( {\bf B} \cdot {\bf k} \big)^2}{4 \pi \rho},
\end{equation}
where ${\bf B}$ is the magnetic field and $\rho$ is the density. The Alfv\'en frequency can also be expressed as 
\begin{equation}
\label{eqn:alven2}
\omega_A^2 = v_A^2 k^2 \mu^2,
\end{equation}
where $v_A$ is the Alfv\'en speed,
\begin{equation}
\label{eqn:valfven}
v_A^2 = \frac{B^2}{4 \pi \rho} \, .
\end{equation}
and $\mu= \cos \theta$ is the angle between the magnetic field and wave vector.

Eqn. \ref{eqn:disp} has two classes of solutions corresponding to each term in parentheses: Alfv\'en waves and magneto-gravity waves. Alfv\'en waves satisfy $\omega^2 = \omega_A^2$ and have wavenumber
\begin{equation}
\label{eqn:alvendisp}
k^2 = \frac{\omega^2}{\mu^2 v_A^2}.
\end{equation}
Alfv\'en waves have fluid velocity perpendicular to the field lines and group velocity $v_g = v_A$ parallel to magnetic field lines. 

Magneto-gravity waves have $\omega^2 = k_\perp^2 N^2/k^2 + \omega_A^2$. A little algebra demonstrates that their wavenumber is
\begin{equation}
\label{eqn:magnetodisp}
k^2 = \frac{\omega^2}{2 v_A^2 \mu^2} \bigg[ 1 \pm \sqrt{1 - \frac{4 \mu^2 v_A^2 N^2 k_\perp^2}{\omega^4}} \bigg].
\end{equation}
The positive and negative roots correspond to the ``slow" and ``fast" magneto-gravity waves, respectively. In the limit of vanishing magnetic field or buoyancy ($v_A \rightarrow 0$ or $N \rightarrow 0$), the slow waves reduce to Alfv\'en waves,
\begin{equation}
\label{eqn:magnetodisp1}
k^2 \simeq \frac{\omega^2}{\mu^2 v_A^2} \, .
\end{equation}
The fast waves reduce to gravity waves,
\begin{equation}
\label{eqn:magnetodisp2}
k^2 \simeq \frac{N^2 k_\perp^2}{\omega^2} \, .
\end{equation}
Gravity waves have fluid velocity nearly perpendicular to the stratification (i.e., nearly horizontal). Their group velocity is primarily horizontal, with
\begin{equation}
\label{eqn:vgravperp}
v_{g,\perp} = \frac{\omega}{k_\perp} \, ,
\end{equation}
but with a small radial component of
\begin{equation}
\label{eqn:vgravr}
v_{g,r} = \frac{\omega^2}{N k_\perp} \, . 
\end{equation}

In the limit of very strong magnetic field or stratification (such that the second term in the square root of Eqn. \ref{eqn:magnetodisp} dominates), the wavenumber obtains a large imaginary component. Therefore, magneto-gravity waves become evanescent in regions of very strong magnetic field. Low frequency waves approaching regions of high field strength can reflect off the stiff field lines, similar to low frequency fluid waves reflecting off a solid boundary. The evanescent skin depth is small, with $H_{\rm ev} \sim \sqrt{v_{A}/(N k_\perp)} \ll H$ when the second term in the square root of Eqn. \ref{eqn:magnetodisp} dominates.

The transition from propagating to evanescent magneto-gravity waves occurs when
\begin{equation}
\label{eqn:magnetogravity}
2 \mu v_A = \frac{\omega^2}{N k_\perp} \, ,
\end{equation}
i.e., when
\begin{equation}
\label{eqn:magnetogravity2}
v_{A,r} \sim v_{g,r} \, .
\end{equation}
Here, we have used $\mu v_A \sim v_{A,r}$ because $k_r \gg k_\perp$ for gravity waves in the WKB limit, and therefore ${\bf B} \cdot {\bf k} \approx B_r k_r$, unless the field is almost completely horizontal. Hence, the \emph{radial} component of the field typically dominates the interaction between the magnetic field and gravity waves. The physical reason for this is that the large horizontal motions and vertical wavenumbers of gravity waves generate large magnetic tension restoring forces by bending radial magnetic field lines. 

Fig. \ref{fig:Prop2} shows wave speeds and wavenumbers corresponding to the propagation diagram in Fig. \ref{fig:Prop}. We note that the Alfv\'en speed is always much less than the sound speed, i.e., the magnetic pressure is much smaller than the gas pressure and the magnetic field has a negligible effect on the background stellar structure. We also note that both Alfv\'en and magneto-gravity waves always have $k \gg 1/H$ and $k \gg 1/r$ in the inner core of our RGB models. Therefore, the WKB analysis used above is justified.

Several previous works (e.g., \cite{Mathis_2012} and references therein) have examined the propagation of magneto-gravity waves in stellar interiors, focusing primarily on the solar tachocline. However, nearly all of these works have considered a purely toroidal (horizontal) magnetic field configuration, because they were motivated by the strong toroidal field thought to exist due to the shear flows in the solar tachocline. Horizontal fields must be stronger by a factor $k_r/k_\perp \sim N/\omega \gg 1$ in order to strongly affect gravity waves. Consequently, these works did not examine the extremely important effect of radial magnetic fields on gravity wave dynamics.

Finally, many papers (such as \cite{Saio_2012} and references therein) have examined the effect of magnetic fields on the acoustic oscillations of rapidly oscillating Ap stars. In this case, the magnetic field strongly affects the acoustic waves only near the surface of the star where the magnetic pressure becomes comparable to the gas pressure. These authors reach similar conclusions to those discussed below: some wave energy can be lost by transmission into Alfv\'en waves, and the geometry of the magnetic field is important. However, the oscillation modes in these stars indicates that observable modes can still exist in the presence of strong magnetic fields, and future studies should further examine possible connections between the physics of oscillating Ap stars and red giants with magnetic cores.

\subsection{Reflection/Transmission}

We define the magneto-gravity radius, $r_{\rm MG}$, as the radius where $\omega \! = \! \omega_{\rm MG}$. At this location, magneto-gravity waves become evanescent and can no longer propagate inward. An incoming wave must either reflect or propagate inward as a pure Alfv\'en wave.

Incoming $\ell \! = \! 1$ magneto-gravity waves can transmit energy into a continuous spectrum \cite{Reese_2004,Levin_2006} of Alfv\'en waves with a broad spectrum of $\ell$ values \cite{Rincon_2003}. Reflected waves will also transfer energy to high $\ell$ waves (for reasons below), and because the location of $r_{\rm MG}$ is a function of latitude since the magnetic field cannot be spherically symmetric. Even in the simplest case of a purely dipolar magnetic field, any resulting oscillation modes will contain a broad spectrum of $\ell$ \cite{Lee_2007,Lee_2010}. In reality, the field will likely have a complex geometry containing both poloidal and toroidal components \cite{Braithwaite_2004,Braithwaite_2006,Duez_2010}, and dipole waves will inevitably scatter into higher $\ell$ waves in the presence of a strong magnetic field.

Wave reflection or transmission at $r_{\rm MG}$ is analogous to the propagation of light between materials of differing refractive indices. In the present case, magneto-gravity waves will likely be reflected due to the high effective refractive index at $r_{\rm MG}$ due to the differing speeds of magneto-gravity and Alfv\'en waves. Just above $r_{\rm MG}$, the group velocity of the incoming magneto-gravity waves is primarily horizontal and is approximately
\begin{equation}
\label{vgroup}
{\bf v}_g  \sim \bigg[ \frac{\omega}{k_\perp} {\bf \hat{n}}_\perp - \frac{\omega^2}{N k_\perp} {\bf \hat{r}} \bigg]
\sim \bigg[ \frac{\omega}{k_\perp} {\bf \hat {n}}_\perp - v_{A,r} {\bf \hat{r}} \bigg] \, .
\end{equation}
Below $r_{\rm MG}$, the group velocity of Alfv\'en waves is $v_A {\bf \hat{B}}$, in the direction of the magnetic field. Thus, although the radial group velocity of the incoming magneto-gravity waves is comparable to that of Alfv\'en waves, their horizontal group velocity is much larger than the Alfv\'en velocity. Except in the case of nearly horizontal fields, coupling to Alfv\'en waves requires a large change in both direction and magnitude of the group velocity. The same is true for the phase velocity. This may cause most of the gravity wave energy to reflect at $r_{\rm MG}$ rather than being transmitted into Alfv\'en waves. 

In the solar atmosphere, an analogous process occurs where magneto-acoustic-gravity waves become magnetically dominated as they propagate upward. In general, the reflection or transmission of the wave depends on the geometry of the magnetic field \cite{Zhugzhda_1984}. Mostly radial fields tend to reflect waves downward at the effective value of $r_{\rm MG}$ in the solar atmosphere \cite{Newington_2009,Newington_2011}. Moreover, the waves are reflected onto the slow branch, i.e., they transition into Alfv\'en waves as they propagate downward. The same process may occur in stellar interiors: ingoing waves will mostly reflect at $r_{\rm MG}$ and will then transition into Alfv\'en waves as they propagate back outward. Sufficiently horizontal fields will allow more wave transmission into Alfv\'en waves in the core, however, stronger fields are required in this case.

\begin{figure}[h!]
\begin{center}
\includegraphics[width=0.9\columnwidth]{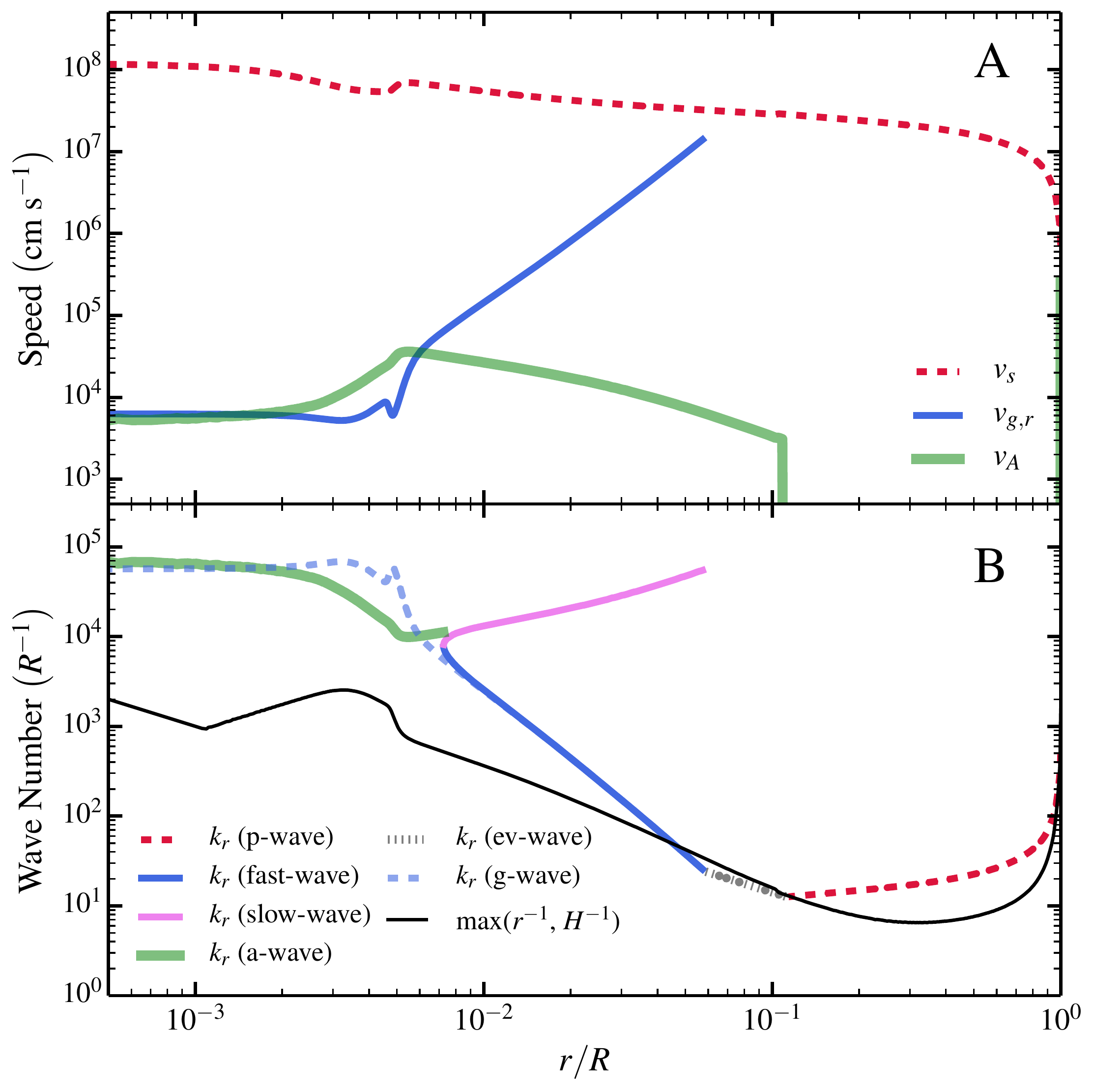}
\caption{\label{fig:Prop2}
Wave properties of the stellar model shown in Figs. \ref{fig:Prop} and \ref{Fig:Struc}. ({\bf A}) The sound speed, $v_s$, gravity wave radial group velocity, $v_{g,r}$ (for $\omega \! = \! 2 \pi \nu_{\rm max}$), and Alfv\'en speed, $v_A$. Waves travel at a group velocity of $v_s$, $v_{g,r}$, and $v_A$ for acoustic, gravity, and Alfv\'en waves, respectively. {\bf (B)} Radial wave number, $k_r$, for a wave with $\omega \! = \! 2 \pi \nu_{\rm max}$, as a function of radius. We have plotted the wavenumbers of acoustic waves (p-waves), fast magneto-gravity waves (fast-waves), slow magneto-gravity waves (slow-waves), Alfv\'en waves (a-waves), and the evanescent part of the wave (ev-wave). The dashed blue line shows the wavenumber for gravity waves (g-waves) in the absence of a magnetic field. The solid black line shows the maximum of $r^{-1}$ or $H^{-1}$ (where $H$ is a pressure scale height).}
\end{center}
\end{figure}

The reflected waves will dissipate much faster than the incident dipole waves, preventing them from ever tunneling back to the surface. Waves reflected back onto the fast branch will have higher $\ell$, shorter wavelengths, and will damp out more quickly than dipole waves. Waves reflected onto the slow branch have wavenumbers orders of magnitude larger than the fast branch of magneto-gravity waves (see Fig. \ref{fig:Prop2}) as they propagate outward into weakly magnetized regions. Therefore, any wave energy reflected into slow magneto-gravity waves will be quickly dissipated via radiative diffusion. 

For perfect wave trapping in the core, purely dipole modes only exist in the envelope, with part of their energy leaking into the core as running magneto-gravity waves. If some wave energy does escape the core, it may leave a signature in the form of mixed magneto-gravity acoustic modes, or by producing magnetic mode splitting, which could be used to constrain the internal magnetic field geometry.

\subsection{Ray Tracing}
\label{ray}

Additional understanding of magneto-gravity waves can be gained using a ray tracing technique. This process allows us to explicitly follow the time evolution of a wave as it propagates into a region of increasing magnetic field. We follow the basic technique outlined in \cite{Ligni_res_2009}. In the case of magneto-gravity waves in the WKB limit, the Hamiltonian describing their equations of motion is
\begin{equation}
\label{eqn:hamiltonian}
H = \omega = \sqrt{ \frac{{\bf k_\perp}^2 N^2}{{\bf k}^2} + ({\bf k} \cdot {\bf v_A})^2 } \, .
\end{equation}
In reality, the Hamiltonian contains additional terms that allow for the existence of pure Alfv\'en waves, although we neglect this subtlety here.

The equations of motion corresponding to the Hamiltonian of Eqn. \ref{eqn:hamiltonian} are
\begin{equation}
\label{eqn:dxdt}
\frac{ d{\bf x}}{dt} = \frac{ \partial H}{\partial {\bf k}} = \frac{N^2}{\omega k} \bigg[ \bigg( 1 - \frac{k_\perp^2}{k^2} \bigg) \frac{{\bf k_\perp}}{k} - \frac{k_\perp^2}{k^2} \frac{k_r {\bf \hat{r}}}{k} \bigg] + \frac{\omega_A}{\omega} {\bf v_A} \, ,
\end{equation}
where $\omega_A = ({\bf k} \cdot {\bf v_A})$, and 
\begin{equation}
\label{eqn:dkdt}
\frac{ d{\bf k}}{dt} = - \frac{ \partial H}{\partial {\bf x}} = - \frac{N}{\omega} \frac{k_\perp^2}{k^2} \nabla N - \frac{\omega_A}{\omega} \nabla \big({\bf k} \cdot {\bf v_A} \big) \, .
\end{equation}

Eqn. \ref{eqn:dxdt} describes the group velocity of the wave, while Eqn. \ref{eqn:dkdt} describes the evolution of its wave vector, which is related to the momentum of the wave. Note that in the absence of a magnetic field in a spherical star, only the radial component of the wave vector changes, and the horizontal component is conserved. This is not surprising because the Hamiltonian is spherically symmetric and thus angular momentum (and hence angular wave vector) is conserved. 

However, in the presence of a magnetic field, the last term of Eqn. \ref{eqn:dkdt} breaks the spherical symmetry. Except in the unphysical case of a purely radial field or a constant field, this term is non-zero, and therefore the angular component of the wave vector must change. At the radius $r_{\rm MG}$ where $v_A \! \sim \! \omega^2/(N k_\perp)$, each term in Eqn. \ref{eqn:dkdt} is the same order of magnitude, assuming $|\nabla B|/B \! \sim \! 1/r$. Therefore, the rate of change in horizontal wavenumber is comparable to the rate of change in radial wavenumber at field strengths near $B_c$. Upon wave reflection or conversion into Alfv\'en waves, the radial wavenumber will generally change by order unity, i.e., the change in radial wavenumber is $|\Delta k_r| \! \sim \! |k_r|$. We therefore expect a correspondingly large change in $k_\perp$, such that $|\Delta k_\perp| \! \sim \! |k_r|$. Hence, dipole waves will generally obtain high multipole moments when they propagate through strongly magnetized regions of the star.

\subsection{Joule Damping}

A gravity wave propagating through a magnetized fluid induces currents which dissipate in a non-perfectly conducting fluid, causing the wave to damp. For gravity waves in the WKB limit which are not strongly altered by magnetic tension forces, the perturbed radial magnetic field is $\delta B \! \approx \! \xi_\perp k_r B$, where $\xi_\perp$ is the horizontal wave displacement. The perturbed current density is $\delta J \! \approx \! c k_r \delta B/(4 \pi)$, where $c$ is the speed of light. The volumetric energy dissipation rate is $\dot{\varepsilon} \! \approx \! (\delta J)^2/\sigma$, where $\sigma$ is the electrical conductivity. The gravity wave energy density is $\varepsilon \! \approx \! \rho \omega^2 \xi_\perp^2$, so the local damping rate is
\begin{equation}
\label{eqn:joule}
\Gamma_B = \frac{\dot{\varepsilon}}{\varepsilon} \approx \frac{\eta B^2 k_r^4}{(4 \pi)^2 \rho \omega^2} \, ,
\end{equation}
where $\eta \! = \! c^2/\sigma$ is the magnetic diffusivity.

The Joule damping rate of Eqn. \ref{eqn:joule} can be compared with the damping rate from radiative diffusion (in the absence of composition gradients), $\Gamma_r \! = \! k_r^2 \kappa$, where $\kappa$ is the thermal diffusivity. The ratio of Joule damping to thermal damping is
\begin{equation}
\label{eqn:jouleratio}
\frac{\Gamma_B}{\Gamma_r} = \frac{\eta}{\kappa} \frac{B^2 k_r^2}{(4 \pi)^2 \rho \omega^2} = \frac{\eta}{\kappa} \frac{l(l+1) B^2 N^2}{(4 \pi)^2 \rho r^2 \omega^4} \, ,
\end{equation}
and the second equality follows from using the gravity wave dispersion relation. The maximum magnetic field possible before Lorentz forces strongly alter gravity waves is $B_c$ (Eqn. \ref{eqn:Bc}), and putting this value into Eqn. \ref{eqn:jouleratio} we find
\begin{equation}
\label{eqn:jouleratio2}
\frac{\Gamma_B}{\Gamma_r} = \frac{1}{16 \pi} \frac{\eta}{\kappa} \, .
\end{equation}
Therefore, for gravity waves, Joule damping cannot exceed thermal damping unless the magnetic diffusivity is significantly larger than the thermal diffusivity. In stellar interiors (and our RGB models), the magnetic diffusivity is typically orders of magnitude smaller than the thermal diffusivity. Therefore Joule damping can safely be ignored. We note that the same result occurs if we use the Alfv\'en wave dispersion relation in Eqn. \ref{eqn:jouleratio}, so Joule damping is also unimportant for Alfv\'en waves.

\subsection{Measurements and Uncertainties} 

Most of the observational data shown in Fig. \ref{fig:moneyplot} were obtained from \cite{Mosser_2012}. The additional stars KIC 8561221 and KIC 9073950 were analyzed using the same methods as \cite{Mosser_2012}. This analysis provided measured values of dipole mode visibility $V^2$,  $\nu_{\rm max}$, $\Delta \nu$, and their associated uncertainties. For KIC9073950, we used the updated KIC $T_{\rm eff}$ \cite{2014ApJS..211....2H} to calculate mass and its uncertainty from scaling relations. For KIC8561221, mass and uncertainties were obtained from \cite{Garcia_2014}. To calculate values of $B_c$ for KIC8561221 and KIC9073950, we interpolated in $\log B_c$ between the tracks shown in Fig. \ref{fig:Bc}, using the measured stellar masses. The uncertainty in $B_c$ was obtained by performing the same interpolation on the upper and lower bounds of the stellar mass.

\begin{table*}
\begin{center}
\begin{tabular}{cccccc}
\hline
Star & $\nu_{\rm max}$ ($\mu$Hz) & $\Delta \nu$ ($\mu$Hz) & $T_{\rm eff}$ (K) & $M$ ($M_\odot$) & $B_{r}$ (G) \\
\hline
KIC8561221 & $490 \pm 24$ & $29.88 \pm 0.80$ & $5245 \pm 60$ & $1.5\pm0.1$ & $ 1.5^{+2.4}_{-0.4} \times 10^7$ \phantom{\huge{X}} \\
\hline
KIC9073950 & $291 \pm 25$ & $20.99 \pm 0.64$ & $5087 \pm 200$ & $1.2\pm0.2$ & $ > 1.3 \times 10^6$ \phantom{\huge{X}} \\
\hline
\end{tabular}
\end{center}
\caption{\label{tab:table} Properties of stars shown in Fig. \ref{fig:Bc}. }
%{\small }
\end{table*}

\subsection{$\varepsilon$ Ophiuchi}

The red giant $\varepsilon$ Ophiuchi, extensively observed with ground-based instruments \cite{De_Ridder_2006} and with the {\it MOST} satellite \cite{Barban_2007}, may also exhibit depressed dipole modes. Its temperature of $\approx 4900 \, {\rm K}$, inferred mass of $1.85 \pm 0.05 \, M_\odot$ and interferometricly measured radius of $10.39 \pm 0.07 \, R_\odot$ \cite{Mazumdar_2009} yield $\nu_{\rm max} \approx 57 \, \mu{\rm Hz}$ and $\Delta \nu \approx 5.5 \, \mu{\rm Hz}$. This is consistent with the interpretation \cite{De_Ridder_2006,Barban_2007,Mazumdar_2009} that many of the peaks in its {\emph MOST} power spectrum belong to a series of radial oscillation modes. However, we agree with \cite{Kallinger_2008} that the most likely explanation for the power spectrum is that it is created by a combination of both radial and non-radial modes.

We speculate that the low amplitude and missing dipole modes can be explained if  $\varepsilon$ Ophiuchi is a depressed dipole mode star. At this stage of evolution, we expect the normalized depressed dipole mode power $V^2$ and lifetime $\tau$ to be roughly half their normal values. The measured lifetimes of $\tau \sim 12 \, {\rm days}$ \cite{Kallinger_2008} are dominated by radial and envelope-dominated quadrupole modes, and are consistent with the usual lifetimes of these modes in red giants at this stage of evolution. A more robust conclusion would require a comparison of measured radial mode line widths to dipole mode line widths, and our scenario would predict that the dipole modes should have lifetimes of $\tau \sim 6 \, {\rm days}$. We suspect that overlapping radial and quadrupole modes may help explain the large line widths found by \cite{Barban_2007}, who considered the peaks to be produced solely by radial modes.

\subsection{MESA Inlist}
\label{inlist}

Here is the inlist used to calculate the stellar evolution models discussed in the paper.
\begin{verbatim}
&star_job            
      change_lnPgas_flag = .true.
      new_lnPgas_flag = .true.
      pgstar_flag = .true.
/ ! end of star_job namelist

&controls      
      !----------------------------------------  MAIN       
      initial_mass = 1.3
      initial_z = 0.02
      use_Type2_opacities = .true.
      Zbase = 0.02        
      !----------------------------------------  WIND
      RGB_wind_scheme = 'Reimers'
      Reimers_wind_eta = 0.5d0 
      RGB_to_AGB_wind_switch = 1d-4
      AGB_wind_scheme = 'Blocker'
      Blocker_wind_eta = 5d0  ! 0.7d0 
      !----------------------------------------  OVERSHOOTING    
      overshoot_f_below_nonburn = 0.018
      overshoot_f_above_burn_h = 0.018
      overshoot_f_above_burn_he = 0.018    
      !----------------------------------------  MISC
      photostep = 100
      profile_interval = 100
      max_num_profile_models = 100
      history_interval = 1
      terminal_cnt = 10
      write_header_frequency = 10
      max_number_backups = 50
      max_number_retries = 100
      max_timestep = 3.15d14  ! in seconds  
      !----------------------------------------  MESH      
      mesh_delta_coeff = 0.8     
      !----------------------------------------  STOP WHEN
      xa_central_lower_limit_species(1) = 'he4'
      xa_central_lower_limit(1) = 0.05    
/ ! end of controls namelist
&pgstar        
/ ! end of pgstar namelist
\end{verbatim}

\end{document}